\documentclass[11pt]{article}
\usepackage{a4}
\usepackage{amsthm}
\usepackage{amssymb}
\usepackage{amsfonts}
\usepackage{amsmath}

\usepackage{tikz}
\usepackage{caption}
\usepackage{subcaption}

\usepackage{color}
\newcommand{\cblue}{\color{blue}}
\newcommand{\cred}{\color{red}}
\newcommand{\cgreen}{\color{green}}

\newcommand {\E}{\mathbb {E}}
\newcommand {\Prob}{\mathbb {P}}
\newcommand{\R}{\mathbb{R}}
\newcommand{\bbF}{\mathbb{F}}
\newcommand{\bbG}{\mathbb{G}}
\newcommand{\bbH}{\mathbb{H}}

\newcommand{\bbN}{\mathbb N}
\newcommand{\sA}{\mathcal A}

\newcommand{\sG}{\mathcal G}
\newcommand{\sH}{\mathcal H}
\newcommand{\sM}{\mathcal M}
\newcommand{\sQ}{\mathcal Q}
\newcommand{\sX}{\mathcal X}

\begin{document}

\title{More on hedging American options under model uncertainty}

\date{\today}
\author{David Hobson \and Anthony Neuberger}

\maketitle

\begin{abstract}
The purpose of this note is to reconcile two different results concerning the model-free upper bound on the price of an American option, given a set of European option prices. Neuberger (2007, `Bounds on the American option') and Hobson and Neuberger (2016, `On the value of being American') argue that the cost of the cheapest super-replicating strategy is equal to the highest model-based price, where we search over all models which price correctly the given European options. Bayraktar, Huang and Zhou (2015, `On hedging American options under model uncertainty', SIAM J. Financial Mathematics) argue that the cost of the cheapest super-replicating strategy can strictly exceed the highest model-based price. We show that the reason for the difference in conclusion is that Bayraktar et al do not search over a rich enough class of models.




\end{abstract}

\section{Overview}
\label{sec:overview}
Our results can be described in a simple discrete-time discrete-space tree and we work in that setting. Let $T \in \bbN$ be a fixed time horizon and let $X=(X_t)_{0 \leq t \leq T}$ be a stock price process. We assume $X_t$ is constrained to lie in $\sX_t$ where $(\sX_t)_{0 \leq t \leq T}$ is a family of potential price levels indexed by time $t$. We work in a discounted universe where all cash values are expressed relative to the bond numeraire and we suppose that the prices of vanilla European call options are given for $X$, so that the one-marginals (but not the joint marginals) of $X$ are specified.

Let $a = \{ a(t,x) \}_{ \{ t \in \{0, 1, \ldots T \}, x \in \sX_t \} }$ be the non-negative payoff function of an American option. Our goal is to consider the cheapest super-replication price $\Psi^a$ of the payoff $a$ and to relate this price to the highest model-based price $\Phi^a$, where in finding the highest model-based price we restrict our search to models which correctly price the vanilla European options. Neuberger~\cite{Neuberger:07} and Hobson and Neuberger~\cite{HobsonNeuberger:16} argue that there is no duality gap, i.e. that $\Phi^a = \Psi^a$. Bayraktar et al~\cite{BayraktarHuangZhou:15} argue that the highest model-based price may be strictly below the cost of the cheapest super-replicating strategy. The aim of this note is to reconcile these contradictory claims.

\section{The example}
\label{sec:eg}

\subsection{Set-up}
\label{ssec:setup}
We proceed largely by studying a simple example. Let $T=2$, let $\sX_0 = \{2\}$, $\sX_1 = \{1,3\}$, $\sX_2 = \{0,2,4\}$. We suppose that a complete set of Arrow-Debreu securities on $X$ are traded, but note that it is sufficient to assume the existence of a single Arrow-Debreu security with maturity 2 (or a single European call with maturity 2 and strike $K \in (0,4)$), as together with the martingale property the price of one non-trivial maturity-2 derivative is sufficient to specify the marginal laws of $X$. Suppose these securities fix the time-2 law of $X$ to be
\begin{equation}
\label{eq:consistency}
 \Prob(X_2 = 0)=\frac{2}{5}; \hspace{10mm} \Prob(X_2 = 2)=\frac{1}{5}; \hspace{10mm} \Prob(X_2 = 4)=\frac{2}{5}.
 \end{equation}
Note, by the martingale property we must have $\Prob(X_1 = 1)=\frac{1}{2}=\Prob(X_1 = 3)$.

Suppose the American option has payoff 1 on the set $\{X_1=1\}$, payoff 8 on the set $\{ X_2 = 4 \}$, and payoff zero otherwise.

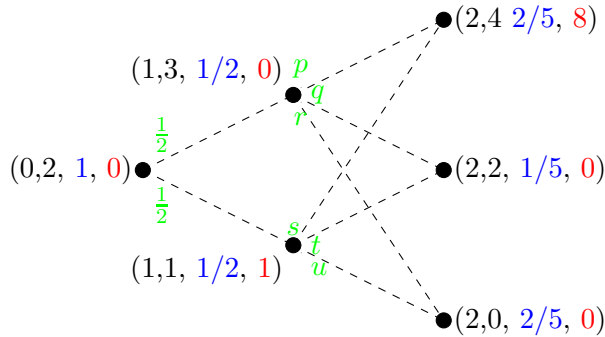
\begin{figure}[!htbp]
\centering
\begin{tikzpicture}
 \draw[dashed] (0,0) -- (2,1) ;
 \draw[dashed] (0,0) -- (2,-1);
 \draw[dashed] (2,1) -- (4,0) ;
 \draw[dashed] (2,1) -- (4,-2);
 \draw[dashed] (2,1) -- (4,2);
 \draw[dashed] (2,-1) -- (4,-2);
 \draw[dashed] (2,-1) -- (4,0);
 \draw[dashed] (2,-1) -- (4,2);
 \draw [fill] (4,2) circle [radius=.1] node[right]{(2,4 {\cblue 2/5}, {\cred 8})};
  \draw [fill] (4,0) circle [radius=.1] node[right]{(2,2, {\cblue 1/5}, {\cred 0})};
  \draw [fill] (4,-2) circle [radius=.1] node[right]{(2,0, {\cblue 2/5}, {\cred 0})};
  \draw [fill] (0,0) circle [radius=.1] node [left] {(0,2, {\cblue 1}, {\cred 0})};
  \draw [fill] (2,1) circle [radius=.1] node [above left] {(1,3, {\cblue 1/2}, {\cred 0})};
  { \draw [fill] (2.1,1.1) circle [radius=0] node [above] {\cgreen $p$};
  \draw [fill] (2.1,1) circle [radius=0] node [right] {\cgreen $q$};
  \draw [fill] (2.1,0.9) circle [radius=0] node [below] {\cgreen $r$}; }
  \draw [fill] (2,-1) circle [radius=.1] node [below left] {(1,1, {\cblue 1/2}, {\cred 1})} ;
  {\draw [fill] (2,-1) circle [radius=0] node [above] {\cgreen $s$};
  \draw [fill] (2.1,-1) circle [radius=0] node [right] {\cgreen $t$};
  \draw [fill] (2.1,-1.1) circle [radius=0] node [below right] {\cgreen $u$}; }
  {\draw [fill] (0,0.1) circle [radius=0] node [above right] {\cgreen $\frac{1}{2}$};
  \draw [fill] (0,-0.1) circle [radius=0] node [below right] {\cgreen $\frac{1}{2}$};}
\end{tikzpicture}%

\caption{{\it The space of possible paths, and the payoff of the American claim}.
The labels at the nodes on the graph consist of a quadruple, the elements of which are time, price level, {\cblue node probability} and {\cred payoff of the American option} respectively. Here node probabilities are equivalent to the prices of Arrow-Debreu securities. {\cgreen Also labelled on the figure by $\frac{1}{2}, \frac{1}{2}$ and $p,q,r,s,t,u$ are conditional probabilities of transitions between nodes.}  }
\label{fig:simpleeg}
\end{figure}

The situation is described in Figure~\ref{fig:simpleeg}.

\subsection{Super-replication}
\label{ssec:sr}
Consider the following super-replicating strategy for this situation:
\begin{quote}
Purchase 1 unit of stock and hold it to maturity or until the exercise of the American option, whichever comes first. In addition, purchase 4 Arrow-Debreu securities with unit payoff on the set $\{ X_2 = 4 \}$, i.e. if the price process is at price level 4 at time 2.
\end{quote}
The cost of this strategy is $2 + 4 \times \frac{2}{5} = \frac{18}{5}$. It is easy to check that this is a super-replicating strategy (for example, if the American option is exercised at $t=1$ when $X_1=1$, then the stock holding is sold for 1 which is sufficient to cover the option obligation, with the possibility of strict super-replication if $X_2=4$). Hence $\Psi^a$, defined to be the cost of the cheapest super-replicating strategy, is at most $\frac{36}{10}$. (We will argue in the final remarks of Section~\ref{ssec:model2} that the strategy above is a cheapest super-replicating strategy and  $\Psi^a=\frac{36}{10}$.) Note the super-replicating strategy of minimal cost need not be unique.

\subsection{The highest model-based price: a first pass.}
\label{ssec:model1}

Let $\Omega_X$ be the set of paths through the nodes and let $\bbF$ be the natural filtration generated by the price process $X$.
Let $\sM_\bbF$ be a space of martingale models for $X$. A martingale model $M \in \sM_\bbF$ can be identified with the set of (conditional) probabilities $(p_{t,i,j})_{\{t = 0,1; i \in \sX_t, j \in \sX_{t+1} \} }$ where
\[ p_{t,i,j} = \Prob^M(X_{t+1}=j| X_{t}=i) \]
and the superscript $M$ indicates that we are taking probabilities under model $M$.
From the requirement that $X$ is a $(\Omega_X, \bbF, \Prob^M)$-martingale we deduce $p_{0,2,j}= \frac{1}{2}$ for $j \in \{1,3\} = \sX_1$, and if we set (see Figure~\ref{fig:simpleeg})
\[ p_{1,3,4}= p;  \hspace{2mm} p_{1,3,2}= q;  \hspace{2mm} p_{1,3,0}= r;  \hspace{12mm} p_{1,1,4}= s;  \hspace{2mm} p_{1,1,2}= t;  \hspace{2mm} p_{1,1,0}= u \]
we find from the martingale condition that ($q = \frac{3-4p}{2}$, $r = p-\frac{1}{2}$, $t = \frac{1-4s}{2}$, $u = \frac{1+2s}{2}$) where $p$ and $s$ are subject to the constraints $\frac{1}{2} \leq p \leq \frac{3}{4}$ and $0 \leq s \leq \frac{1}{4}$.
We may parameterise martingale models $M \in \sM_{\bbF}$ by $p$ and $s$ and write $M^{p,s}$ for the model for which
$p_{1,3,4}= p$ and $p_{1,1,4}= s$.

For $M^{p,s} \in \sM_\bbF$ consider the model-based price
$ \phi^a(M^{p,s}) = \sup_\tau \E[ a(\tau, X_\tau)]$. By considering the optimal exercise decisions in the possible states at time $1$ we find
\[ \phi^a(M^{p,s}) = \frac{1}{2}\max \{ 1 , 8s \} + \frac{1}{2} 8p = 4(p+s) + \frac{(1-8s)^+}{2}. \]

Let $\sQ_\bbF \subseteq \sM_\bbF$ be the space of martingale models of this form for which the law of $X_2$ matches the law specified by the Arrow-Debreu securities. Then for $M \in \sQ_\bbF$ we must have $\frac{p+s}{2} = \frac{2}{5}$. For $M^{p,s}\in \sQ_\bbF$,
\[ \phi^a(M^{p,s}) = \frac{16}{5} + \frac{(1-8s)^+}{2}. \]
This is maximised over feasible $p,s$ at $p=\frac{3}{4}, s=\frac{1}{20}$, and we find
\[ \Phi^a_\bbF := \sup_{M^{p,s} \in \sQ_\bbF} \phi^a(M^{p,s}) = \frac{35}{10} \]
In particular, as noted by Bayraktar et al~\cite{BayraktarHuangZhou:15} in a slightly more complicated example, $\Phi^a_\bbF < \Psi^a$.

\subsection{The highest model-based price: a second pass.}
\label{ssec:model2}
Since $\Phi^a_\bbF < \Psi^a$ this raises the immediate question: is there a duality gap? In fact the answer is no, and the strict inequality $\Phi^a_\bbF < \Psi^a$ arises from the fact that the set of models considered in Section~\ref{ssec:model1} and by Bayraktar et al~\cite{BayraktarHuangZhou:15} is insufficiently rich. In their analysis Bayraktar et al~\cite{BayraktarHuangZhou:15} closely follow the set-up in Bouchard and Nutz~\cite{BouchardNutz:15}. But, one of the implict assumptions of Bouchard and Nutz is that it is sufficient to work with the canonical process and the natural filtration. To quote Acciaio et al~\cite{AcciaioBeiglbockPenknerSchachermayer:13}: `The obvious approach consists in considering as admissible martingale measures, all probability measures in which the co-ordinate process is a martingale in its own filtration.' This is fine in the context of finding super-hedges and robust bounds on the prices of path-dependent (European-style) payoffs (as is the case in \cite{AcciaioBeiglbockPenknerSchachermayer:13} and \cite{BouchardNutz:15}), but it is not adequate for American options. With an American option we have to consider the scenario in which information on the dynamics of the underlying arrives after the initiation of the option. Such information may affect the exercise decision of the agent, and therefore the value of the American option.

Consider $\bar{M}$ which is a mixture of the models $M^{\frac{3}{4},0}$ and $M^{\frac{3}{4},\frac{1}{4}}$ such that weight $\frac{4}{5}$ is placed on $M^{\frac{3}{4},0}$ and weight $\frac{1}{5}$ is placed on $M^{\frac{3}{4},\frac{1}{4}}$. Individually, neither $M^{\frac{3}{4},0}$ nor $M^{\frac{3}{4},\frac{1}{4}}$ is an element of $\sQ_\bbF$, but for the mixture model
\[ \Prob^{\bar{M}}(X_2 = 4) =
\frac{4}{5} \Prob^{M^{\frac{3}{4},0}}(X_2 = 4) +
\frac{1}{5} \Prob^{M^{\frac{3}{4},\frac{1}{4}}}(X_2 = 4) = \frac{4}{5} \times \frac{3}{8} + \frac{1}{5} \times
\left[ \frac{3}{8} + \frac{1}{8} \right] = \frac{2}{5} \]
so that $\bar{M}$ is consistent with the given Arrow-Debreu prices. If the option holder can learn which of $M^{\frac{3}{4},0}$ or $M^{\frac{3}{4},\frac{1}{4}}$
describes the price dynamics before deciding whether to exercise the American option at $t=1$, then he can take advantage of this information to refine his stopping rule and obtain a higher expected payoff.

In order to describe the idea of learning about the price dynamics we need to expand our concept of a model beyond the notion that it is a probability measure on $(\Omega_X, \bbF)$. In the context of robust pricing we are given the 1-marginal laws of the price process but we are not given the underlying probability space.  Therefore, we define a model $M$ to be a filtered probability space $(\Omega^M, \sG^M, \bbG^M = (\sG^M_{t})_{t}, \Prob^M)$ supporting the stochastic process $X^M$ (i.e. $X^M$ is $\bbG^M$-adapted) where $X^M$ is the price process. A martingale model is a model $(\Omega^M, \sG^M, \bbG^M = (\sG^M_{t})_{t}, \Prob^M)$ such that $X^M$ is a $(\Omega^M,\bbG^M,\Prob^M)$-martingale, and a consistent model is a martingale model such that the law of $X^M$ under $\Prob^M$ is consistent with the given European option prices, or in our case such that the probabilities $\Prob^M(X^M_2 = j)$ agree with those given in \eqref{eq:consistency}. The crucial point is that the sample space need not just be the set of paths through the nodes of the tree (or just be in 1-1 correspondence with paths in the tree), and the filtration can be strictly larger than the natural filtration of the price process.

Let $\sM$ be the space of all martingale models and let $\sQ$ be the space of all consistent models. Then $\sM_{\bbF} \subseteq \sM$ and $\sQ_{\bbF} \subseteq \sQ$.


Now we wish to describe a model which attains the highest consistent model based price for the American option.
Let $\Omega^* = \Omega_X \times \{0,1 \} = \{(\omega_X,y) \}$ where $\omega_X$ is a path through the tree, i.e. $\omega_X = (x_0,x_1,x_2)$ with $x_0=2$, $x_1 \in \{1,3\} =\sX_1$ and $x_2 \in \{0,2,4\} = \sX_2$. Let $X=(X_t)_{t=0,1,2}$ be the first coordinate process, so that $X_j(\omega_X,y) = x_j$ and
let $Y$ be a Bernoulli random variable such that $Y(\omega_0,y) = y$. Finally, let $\sH_0$ be trivial; for $t=1,2$ let $\sH_t = \sigma(Y, X_s; s \leq t)$; let $\bbH = (\sH_t)_{0 \leq t \leq T}$.

Let $M^*$ be the model $M^*=(\Omega^*, \sH_2, \bbH, \Prob^{*})$ where $X_1$ and $Y$ are independent under $\Prob^{*}$, $\Prob^{*}(Y=0) = \frac{4}{5} = 1 - \Prob^{*}(Y=1)$, and $\Prob^{*}(X_1 = 3) = \Prob^{*}(X_1=1) = \frac{1}{2}$
together with
\begin{eqnarray*}
\lefteqn{\Prob^{*}(X_2 = k| X_1 = j, Y) =} \\
&& I_{ \{ Y=0\} } \Prob^{M^{\frac{3}{4},0}}(X_2 = k| X_1 = j) + I_{ \{ Y=1\} } \Prob^{M^{\frac{3}{4},\frac{1}{4}}}(X_2 = k| X_1 = j) .
\end{eqnarray*}
In the model $M^*$ the option holder learns whether the dynamics are governed by $M^{\frac{3}{4},0}$ or $M^{\frac{3}{4},\frac{1}{4}}$ at time $1$ in time to inform his decision about whether to exercise the American option. 
Conditional on $X_1=1$ and $Y=0$ so that the transition probabilities are as in $M^{\frac{3}{4},0}$ it is optimal to exercise at $t=1$, and the payoff of the American option is 1. Conditional on $X_1=1$ and $Y=1$ so that the transition probabilities are as in $M^{\frac{3}{4},\frac{1}{4}}$ it is optimal to not exercise at $t=1$ and the expected payoff of the option is $2$. Conditional on $X_1=3$ it is always optimal to wait at $t=1$ and the expected payoff of the option is 6. The unconditional expected payoff of the option is
\[ \frac{1}{2} \times \frac{4}{5} \times 1 +  \frac{1}{2} \times \frac{1}{5} \times 2 + \frac{1}{2} \times 6 = \frac{18}{5} \]
In particular the model-based price of the American option $\phi^a(M^*)$ is $\frac{18}{5}$.

We have
\[ \frac{18}{5} \geq \Psi^a \geq \Phi^a := \sup_{M \in \sM} \phi^a(M) \geq \phi^a(M^*) = \frac{18}{5} \]
where the first inequality follows from the existence of the super-replicating strategy at cost $\frac{18}{5}$ given in Section~\ref{ssec:sr}, the second inequality follows from weak duality, and the third inequality follows from the fact that
$M^*=(\Omega^*, \sH_2, \bbH, \Prob^{*})$ is a consistent model. Hence there is equality throughout and $\Psi^a = \Phi^a$ i.e. the cost of the cheapest super-replicating strategy is equal to the highest model-based price. Further, as advertised in Section~\ref{ssec:sr} the strategy given therein is a cheapest super-replicating strategy, and the model $(\Omega^*, \sH_2, \bbH, \Prob^{*})$  gives the highest model based price.


\subsection{Discussion}
\label{ssec:discussion}
As argued in \cite{HobsonNeuberger:16} the full value of an American option arises from the ability of the option holder to adjust their strategy in the light of new information. This is especially true in cases of event risk. If the dynamics of a price depend on an event at some future time within the life of the American option, then the option holder may adjust his exercise strategy in the light of the outcome of this event. Excluding this possibility, by, for example, working in the natural filtration, reduces the potential value of the American option.

Under our notion of a model, the set of all martingale models and the set of all consistent models are both vast. One of the contributions of Hobson and Neuberger~\cite{HobsonNeuberger:16} is to prove that it is possible to restrict attention to a much smaller class of models which are generated by augmenting the price process with a second process called the regime process. The search for the highest model-based price consistent with a set of European options can be reduced to a search over this class.

Whilst we were finalising this note we became aware of a paper by Deng and Tan~\cite{DengTan:16} who also consider the problem of duality for American options in the robust context. They work in  discrete-time, but allow for a continuum of European option prices and prove a duality theorem to show there is no duality gap for a wide class of American-payoffs. Many of their results mirror those of \cite{HobsonNeuberger:16}, but in a more general setting. In addition, they introduce the notions of the weak formulation (i.e. $\sup_{M \in \sM} \phi^a(M)$) and strong formulation ($\sup_{M \in \sM_{\bbF}} \phi^a(M)$) and show these typically lead to different price bounds. In the financial context the starting point is European option prices, rather than a probability space or set of probability spaces, so our contention is that it is the weak formulation which is appropriate for model-free pricing.



Bayraktar et al~\cite[Equation (3.4)]{BayraktarHuangZhou:15} give an alternative representation of the highest model price as (simplified to the setting of our example)
\[  \tilde{\Phi}^a_{\bbF} = \inf_{\beta \in \R}  \sup_{M^{p,s} \in \sM_{\bbF}} \sup_\tau \E^{M^{p,s}} \left[a(\tau, X_\tau) - \beta \left( I_{\{ X_2 = 4 \} } - \frac{2}{5} \right) \right] \]
where $\tau$ is a $\bbF$-stopping time.
We then find, with $\sA = \{ (p,s): \frac{1}{2} \leq p \leq \frac{3}{4}, 0 \leq s \leq \frac{1}{4} \}$
\begin{eqnarray*}
\tilde{\Phi}^a_{\bbF} & = & \inf_\beta \sup_{(p,s) \in \sA} \left[ \phi^a(M^{p,s}) - \beta \left(\frac{p+s}{2} - \frac{2}{5} \right) \right] \\
& = & \inf_\beta \max \left\{ \frac{5}{2} + \frac{3}{20} \beta ; \frac{5}{2} + \frac{3}{20} \beta;\frac{5}{2} + \frac{3}{20} \beta; \frac{5}{2} + \frac{3}{20} \beta \right\}
\end{eqnarray*}
where we use the convexity of the objective function to reduce the optimisation over $\sA$ to an optimisation over the corners of $\sA$. We find that $\beta=4$ and that $\tilde{\Phi}^a_{\bbF}=\frac{36}{10} = \Psi^a$. The maximum is attained at $\beta = 4$ for $(p,s)=(\frac{3}{4}, 0)$ and $(p,s)=(\frac{3}{4}, \frac{1}{4})$, so that this approach is picking out the models in $\sM_{\bbF}$ which from constituents of the mixture model in Section~\ref{ssec:model2}.

Bayraktar et al~\cite{BayraktarHuangZhou:15} also prove results for the subhedging problem and the lowest model price. They show \cite[Theorem 2.1]{BayraktarHuangZhou:15} that the sub-hedging price is equal to the the lowest price under a consistent model, even when attention is restricted to the class $\sM_{\bbF}$.

\section{Conclusions}
\label{sec:conc}
Neuberger~\cite{Neuberger:07} and Hobson and Neuberger~\cite{HobsonNeuberger:16} show that in the setting of American options it is not sufficient to restrict attention to models in which the sample space is the set of candidate paths of the price process and the filtration is the natural filtration of the price process. We use this fact to explain the apparent contradiction between the results of \cite{Neuberger:07,HobsonNeuberger:16} and Bayraktar et al~\cite{BayraktarHuangZhou:15}. Bayraktar et al closely follow the work of Bouchard and Nutz~\cite{BouchardNutz:15} who consider European-style path dependent claims, and who work with the canonical process and the natural filtration. For this reason when Bayraktar et al calculate a highest model price, the space of models they consider is not sufficiently rich as to capture the full value of an American option.

\end{document}